# Deep Learning-Based Attenuation and Scatter Correction of Brain $^{18}$F-FDG PET Images in the Image Domain

Reza Jahangir[1], Alireza Kamali-Asl[1], and Hossein Arabi[2]

[1]Department of Medical Radiation Engineering, Shahid Beheshti University, Tehran, Iran

[2]Division of Nuclear Medicine and Molecular Imaging, Department of Medical Imaging, Geneva University Hospital, CH-1211 Geneva 4, Switzerland

**Abstract**

*Purpose:* Attenuation and scatter correction (AC) is crucial for quantitative Positron Emission Tomography (PET) imaging. Recently, direct application of AC in the image domain using deep learning approaches has been proposed for the hybrid PET/MR and dedicated PET systems that lack accompanying transmission or anatomical imaging. This study set out to investigate deep learning-based AC in the image domain using different input settings.

*Methods:* Three different deep learning methods were developed for the direct application of AC in the image domain: 1) use of PET non-attenuation corrected images as the input (PET-nonAC), 2) use of PET attenuation corrected images with a simple 2-class AC map (composed of background air and the soft tissue) obtained from PET-nonAC [PET Segmented-based AC (PET-SegAC)] images, and 3) use of both PET-nonAC and PET-SegAC images in a dual-channel mode to estimate the reference PET attenuation corrected images with CT images (PET-CTAC). Since a simple 2-class AC map can be easily obtained from PET-nonAC images, this study assessed the added value of incorporating PET-SegAC images into direct AC in the image domain. A dataset of 80 brain 18F-fluorodeoxyglucose PET/CT images was used for the training and evaluation of the different models using a 4-fold cross-validation scheme. The voxel-wise and region-wise accuracy of the models were examined by measuring the standard uptake value (SUV) errors in different regions of the brain.

*Results:* The dual-channel model resulted in an overall 0.156±0.07 SUV root mean square error (RMSE) over the whole head region while the deep learning models with only PET-nonAC and PET-SegAC inputs yielded RMSEs of 0.214±0.07 and 0.189±0.14, respectively. The region-wise analysis of the radiotracer in different brain regions revealed an SUV bias of 0.25%±4.72% for the dual-channel model and SUV biases of 1.42%±5.08% and 0.61%±5.19% for the models with only PET-nonAC or PET-SegAC input images, respectively. The PET-SegAC images, representing a bottom line for clinically tolerable errors, exhibited an SUV bias of -10.46%±5.36%. Altogether, the dual-channel model relying on both PET-SegAC and PET-nonAC images exhibited superior performance over the other models.

*Conclusion*: Since a 2-class AC map can be easily generated from PET-nonAC images, this study investigated the added value of incorporating PET-SegAC images into a deep learning-based direct AC in the image domain. Overall, the dual-channel deep learning model, which relies on both PET-nonAC and PET-SegAC images, would offer a more accurate AC modeling, compared to the models using only PET-nonAC or PET-SegAC images.

## 1. Introduction

Positron Emission Tomography (PET) scan, as a non-invasive functional imaging modality, provides three-dimensional distribution maps of radioactive tracers at the molecular and cellular level enabling the *in vivo* estimation of chemical and biological processes of the organs/tissues. Most neuroimaging PET studies employ 18F-fluorodeoxyglucose (FDG) radiotracer in brain imaging for the diagnosis of neurodegenerative diseases owing to its close correlation with a certain type of metabolism (as a proxy of glucose). In hybrid PET/computed tomography (PET/CT) scanners, when combining the PET imaging system with structural CT modality, anatomical CT images are employed to model/compensate for the underlying physical degrading factors, such as attenuated and scattered photons in PET imaging. Anatomical CT imaging readily provides attenuation coefficient maps for attenuation and scatter correction (AC) in PET imaging; however, in hybrid PET/magnetic resonance imaging (PET/MRI), the generation of patient-specific attenuation maps from MR images is a major challenge for quantitative PET imaging [1]. This stems from the fact that MR intensity reflects the proton density of mediums/tissues rather than electron density, which is required in PET imaging at the photon energy of 511 keV. In addition to the hybrid PET/MRI, the task of AC in dedicated PET systems, designed and fabricated for brain and pelvis imaging, is the major barrier to accurate quantitative PET imaging [2]. Though these dedicated PET scanners provide a relatively higher spatial resolution and sensitivity, compared to the conventional PET/CT or PET/MR systems, they are not normally combined with any anatomical/structural imaging modalities; therefore, the generation of attenuation maps or accounting for attenuated/scattered photons is not a trivial task [3].

The conventional AC map generation strategies in PET/MR imaging are conducted through the bulk segmentation of the MR image into several major tissue types (such as the soft tissue, fat, bone, lung, and air cavities) [4-6] or atlas registration, which relies on a dataset of body/head templates to identify the underlying anatomical structures [7-10]. The MR-assisted joint reconstruction of the attenuation and activity maps is another solution for attenuation correction in PET/MR imaging [11, 12]. Since the advent of deep learning algorithms, a vast number of approaches have been introduced/proposed for the generation of synthetic CTs from MR images [13-17], estimation of the attenuation correction information from PET emission data [18, 19], or applying attenuation correction on PET images using hybrid methods, such as combining deep learning, joint attenuation, and activity map reconstruction [20].

In addition to these approaches, direct application of the AC on non-attenuation corrected PET images (PET-nonAC) without using anatomical information has been introduced as a feasible technique for the task of AC in dedicated PET scanners that lack transmission imaging [3, 21-24]. In this approach, PET-nonAC images are regarded as the input to a deep learning network to predict the attenuation and scatter corrected PET images, considering the CT-based AC (PET-CTAC) as the reference in an end-to-end image translation fashion. In this regard, a simple water-equivalent PET AC map can be simply generated by delineating the body contour from PET-nonAC images and assigning predefined attenuation coefficients (equivalent to the soft tissue) to the voxels inside the body contour, as well as air attenuation coefficients to the voxels outside the contour. The generation of such AC maps does not require anatomical imaging, and they can be easily generated from PET-nonAC images in case of Time-of-Flight (TOF) imaging owning to the high signal-to-noise ratio and strong contrast at the boundary of the body. This approximate AC map can be used for preliminary PET attenuation correction, and then, the resulting Segmentation-based attenuation correction PET images (PET-SegAC) can be fed into a deep learning method to predict accurate PET-CTAC images. Since these PET images are partly/approximately corrected for attenuation and scatter, the prediction of accurate PET-CTAC images might be less challenging and/or more robust for deep learning approaches.

This study aimed to investigate the impact of using an intermediate AC on PET images for the direct AC in the image domain through the assessment of different scenarios based on a deep learning approach. These scenarios include: 1) direct PET-CTAC estimation from PET-nonAC images, 2) PET-CTAC estimation from PET-SegAC images, and 3) PET-CTAC estimation from a dual-channel deep learning approach with both PET-Seg and PET-nonAC images as input channels.

## 2. Materials and Methods

### 2.1. Data Acquisition

A dataset of 80 patients undergone TOF brain PET/CT scans was retrospectively exploited for the evaluation of direct AC in the image domain. The Ethics Committee of Geneva University Hospitals approved the study protocol, and informed consent was obtained from all patients. Patients included in this study were 44 males and 36 females with a mean age of 66±11 years and a weight of 73±18 kg. The PET/CT scans were performed on a Biograph mCT scanner (Siemens Healthcare, Erlangen, Germany) for 20 min in a single bed position. For the task of PET attenuation correction, a low-dose CT image was acquired with the current of 20 mAs, the speed of 0.3 sec/rotation, the energy of 120 kVp, and the voxel size of 0.9×0.9×2.5 mm$^3$. The PET acquisitions started 33±5 min after the injection of 206±13 MBq dose of $^{18}$F-FDG.

### 2.2. Data Preparation

Direct application of the AC on PET images in the image domain was investigated for three different scenarios: 1) use of PET-nonAC images as the input, 2) use of PET-SegAC images as the input, and 3) use of both PET-nonAC and PET-SegAC images in a dual-input mode to estimate the reference PET-CTAC images. For the training of the deep learning model, a dataset consisting of PET-nonAC, PET-SegAC, and

PET-CTAC images was created for each patient. To generate the PET-CTAC (reference image) and PET-nonAC images, the Siemens e7 tool was employed to reconstruct the PET raw data with and without CT-based attenuation map using TOF ordinary Poisson ordered subsets-expectation maximization with 5 iterations and 21 subsets. To generate the PET-SegAC images, a 2-class attenuation map was created from the classification of the TOF reconstructed PET-nonAC images into the background air and head region. Attenuation coefficients of 0.0 cm$^{-1}$ ($\approx$-1000 HU) and 0.1 cm$^{-1}$ ($\approx$0 HU) were assigned to the voxels within the background air and head region, respectively. The reconstruction of the PET raw data was repeated with the aforementioned settings and using the segmentation-based AC map to generate PET-SegAC images. The reconstruction of PET-CTAC and PET-SegAC images involved scatter correction using the single scatter simulation method followed by the tail fitting for the estimation of the scatter scaling factor. The entire PET images were reconstructed in a matrix with 200×200×109 voxels and a voxel size of 2×2×2 mm followed by Gaussian denoising with 2 mm FWHM.

For the training of the deep learning model, the intensity of the PET images should be normalized to a certain common range to facilitate/harmonize the process of feature extraction. To this end, the voxel values of PET-CTAC, PET-SegAC, and PET-nonAC images were in the first step converted to the standard uptake value (SUV). Afterward, an empirical fixed value of 9 was used to scale down the intensities of PET-CTAC and PET-SegAC images. Likewise, PET-nonAC images were normalized by an empirical factor of 3. In an effort to diminish the computation burden, the background air was cut off from PET images leading to a final matrix size of 128×128×105 voxels.

### 2.3. Deep Learning Architecture

For the implementation of the direct AC, the Niftynet platform was exploited, which is an open-source pipeline for the realization of deep learning algorithms. This platform provides modular convolutional neural network facilities for common medical image analysis [25]. The Niftynet platform was built in Python (version 3.6) using the TensorFlow (version 1.12) libraries. For the implementation of the direct PET-CTAC estimation from PET-nonAC or PET-SegAC images, a high resolution residual neural network [26], referred to as HighResNet in the Niftynet, was retrieved and reconfigured. This neural network consists of 20 residual layers which extract/process the image features at different levels (scales) while maintaining the spatial resolution of the input image by applying dilated convolutional kernels. The first seven layers use the convolution kernels of 3×3×3-voxel to extract low-level image features, such as the edges. These convolutional kernels are dilated by factors of 2 and 4 in the following seven and six layers to extract medium- and low-level image features, respectively. A residual connection is then established to link every two convolutional layers, wherein an element-wise rectified linear unit (ReLU) and a batch normalization are connected to the convolutional layers within the residual blocks.

### 2.4. Implementation Details

As mentioned before, three scenarios were followed for the implementation of the direct AC in the image domain: estimation of PET-CTAC from PET-nonAC images, PET-SegAC images, and both PET-nonAC and PET-SegAC (using a dual-channel input) images. A 4-fold cross-validation scheme was followed for the implementation of each scenario considering that the 80 subjects were selected at each fold in the dataset, 60 training, and 20 test subjects.

The residual neural network implemented in the Niftynet platform was trained using a spatial window with the size of 128×128, the Adam optimizer, the learning rate of 0.001, the L2norm loss function, the decay of 0.0001, the batch size of 30, and the sample per volume of 1. The training and inference of these three scenarios were performed in a 2-dimensional mode. Training of the models was performed in 20 epochs taking approximately 10 h to reach the plateau of the training loss function. The models training and inference were carried out on the NVIDIA GTX 1060 with 6 GB of GPU memory and the Linux Ubuntu 18.04 LTS operating system. No transfer learning or pre-trained model was used within the training of these models. The inference of the synthetic PET-CTAC images took less than 10 sec for each patient (one-bed position, whole-brain study).

### 2.4. Evaluation Strategy

The performance of different scenarios of the direct AC was assessed against the CT-based AC considered as the reference. Moreover, PET-SegAC images were included in this evaluation, which provides a bottom line for clinically tolerable errors. The 2-class SegAC obtained from the classification of the TOF PET-nonAC images is regarded as a proxy for the MRI-guided attenuation map generation implemented in the commercial PET-MR scanners [27].

Regarding PET-CTAC images as the reference, the peak signal-to-noise ratio (PSNR), structural similarity index (SSIM), root mean square (RMSE), absolute mean error (MAE), and mean error (ME) were calculated for different PET images using Eqs. 1 to 5, respectively. The synthetic PET-CTAC images estimated by the deep learning models from PET-nonAC images, PET-SegAC images, and both PET-nonAC and PET-SegAC images (dual-channel input) are referred to as PET-DLnon, PET-DLSeg, and PET-DLDC, respectively.

$$PSNR(db) = 20 \log_{10} \frac{\max(PET_{Ref},\ PET_{ASC})}{\sqrt{MSE(PET_{ASC}, PET_{Ref})}} \qquad (1)$$

$$SSIM = \frac{(2\mu_{Ref}\mu_{ASC}+c_1)(2\sigma_{Ref,ASC}+c_2)}{(\mu_{ASC}^2+\mu_{Ref}^2+c_1)(\sigma_{ASC}^2+\sigma_{Ref}^2+c_2)} \quad (2)$$

$$RMSE = \sqrt{\frac{1}{V}\sum_{k=1}^{V}(PET_{ASC}(k)-PET_{Ref}(k))^2} \quad (3)$$

$$MAE = \frac{1}{V}\sum_{k=1}^{V}|PET_{ASC}(k)-PET_{Ref}(k)| \quad (4)$$

$$ME = \frac{1}{V}\sum_{k=1}^{V}(PET_{ASC}(k)-PET_{Ref}(k)) \quad (5)$$

Here, $PET_{ASC}$ represents either PET-SegAC or the deep learning-based synthesized PET images, $PET_{Ref}$ stands for the reference PET-CTAC images. $V$ is the total number of non-zero voxels in the images and $k$ is the voxel index. $Max$ ($PET_{Ref}$, $PET_{ASC}$) returns the maximum value of either $PET_{Ref}$ or $PET_{ASC}$ images, and $MSE$ calculates the mean squared error between the reference and predicted images. The $\mu_{Ref}$ and $\mu_{ASC}$ in Eq. (2) indicate the mean of reference and predicted PET images, respectively, and $\sigma_{Ref,ASC}$ represents the covariance of $\sigma_{ASC}^2$ and $\sigma_{Ref}^2$, which in turn denotes the variance of reference and predicted PET images, respectively. $c_1$ and $c_2$ ($c1=0.01$ and $c2=0.02$) are two variables to stabilize the division with a zero or a very small denominator.

For the region-based analysis of the $^{18}$F-FDG uptakes, the activity concentration was measured in 70 brain regions using the digital atlas of the human brain and automated anatomical labels [28] implemented in Pmode processing platform (version 3.8) [29]. Given the 70 anatomical regions of the brain, the absolute mean bias (ARB%) and relative mean bias (RB%) were calculated for each region across the 80 patients using Eqs. (6) and (7), respectively. Given the activity concentration in the 70 brain regions for different PET images, the Bland-Altman plot of uptake differences with the reference PET-CTAC image in SUV was plotted for different PET images.

$$ARB_{region}(\%) = \left|\frac{(PET_{ASC})_{region}-(PET_{ref})_{region}}{(PET_{ref})_{region}}\right| \times 100\% \quad (6)$$

$$RB_{region}(\%) = \frac{(PET_{ASC})_{region}-(PET_{ref})_{region}}{(PET_{ref})_{region}} \times 100\% \quad (7)$$

Moreover, radiomic features of the 70 anatomical regions of the brain were extracted for all patients and different synthetic PET images using the LIFEx freeware radiomic feature calculation [30]. For each brain region, 28 radiomic features were extracted, including SUV, intensity, Grey-Level Zone Length Matrix, Grey-Level Run Length Matrix, and Grey-Level Co-occurrence Matrix. Detailed radiomic features information is summarized in Supplemental Table 1.

For the voxel-level analysis, a joint histogram evaluation was carried out to illustrate the voxel-wise correlation between the radiotracer activity concentration in PET-CTAC images and each synthetic PET image, including PET-DLnon, PET-DLSeg, PET-DLDC, as well as PET-SegAC.

To investigate the statistical significance of the errors and differences between different models, a paired t-test analysis was conducted for different metrics wherein a P-value of less than 0.05 was considered statistically significant.

## 3. Results

The quantitative analysis of the PET-DLNAC, PET-DLSeg, PET-DLDC, as well as PET-SegAC images for the entire head region, is summarized in Table 1 wherein the mean and standard deviation of the ME, MAE, RMSE, PSNR, and SSIM metrics are reported for all 80 patients. The PET-SegAC images are included to provide a bottom line for clinically tolerable errors. Estimation of the PET-CTAC using the dual-channel deep learning approach with PET-nonAC and PET-SegAC images (referred to as PET-DLCD) as the input exhibited superior accuracy over single-channel deep learning approaches taking PET-nonAC and PET-SegAC images as the input (referred to as PET-DLNAC and PET-DLSeg, respectively). However, all of these three approaches outperformed the PET-SegAC method. Differences between these AC methods are statistically significant as shown in Table 2, particularly for the MAE and RMSE metrics.

**Table 1**. Quantitative evaluation of the PET-DLNAC, PET-DLSeg, PET-DLDC, and PET-SegAC images within the entire head region for 80 patients using PET-CTAC as the reference.

|  | ME (SUV) | MAE (SUV) | RMSE (SUV) | PSNR (dB) | SSIM |
|---|---|---|---|---|---|
| **PET-DLNAC** | $0.004 \pm 0.09$ | $0.152 \pm 0.05$ | $0.214 \pm 0.07$ | $19.08 \pm 3.18$ | $0.987 \pm 0.01$ |
| **PET-DLSeg** | $-0.008 \pm 0.07$ | $0.128 \pm 0.08$ | $0.189 \pm 0.14$ | $20.98 \pm 3.68$ | $0.990 \pm 0.01$ |
| **PET-DLDC** | $0.002 \pm 0.06$ | $0.123 \pm 0.05$ | $0.156 \pm 0.07$ | $21.48 \pm 5.35$ | $0.992 \pm 0.01$ |
| **PET-SegAC** | $-0.461 \pm 0.08$ | $0.483 \pm 0.08$ | $0.715 \pm 0.14$ | $17.24 \pm 1.77$ | $0.969 \pm 0.01$ |

**Table 2**. P-values calculated between different AC methods for the quantitative results reported in Table 1.

|  | ME (SUV) | MAE (SUV) | RMSE (SUV) | PSNR (dB) | SSIM |
|---|---|---|---|---|---|
| **PET-DLNAC vs. PET-DLSeg** | 0.44 | 0.01 | 0.02 | <0.01 | 0.01 |
| **PET-DLNAC vs. PET-DLDC** | 0.04 | <0.01 | <0.01 | <0.01 | <0.01 |
| **PET-DLNAC vs. PET-SegAC** | <0.01 | <0.01 | <0.01 | <0.01 | <0.01 |
| **PET-DLDC vs. PET-DLSeg** | 0.04 | 0.03 | 0.03 | 0.01 | 0.01 |
| **PET-DLSeg vs. PET-SegAC** | <0.01 | <0.01 | <0.01 | <0.01 | <0.01 |
| **PET-DLDC vs. PET-SegAC** | <0.01 | <0.01 | <0.01 | <0.01 | <0.01 |

Figure 1 shows representative views of the PET images corrected for attenuation using different AC methods, as well as the reference PET-CTAC images together with the corresponding difference bias maps.

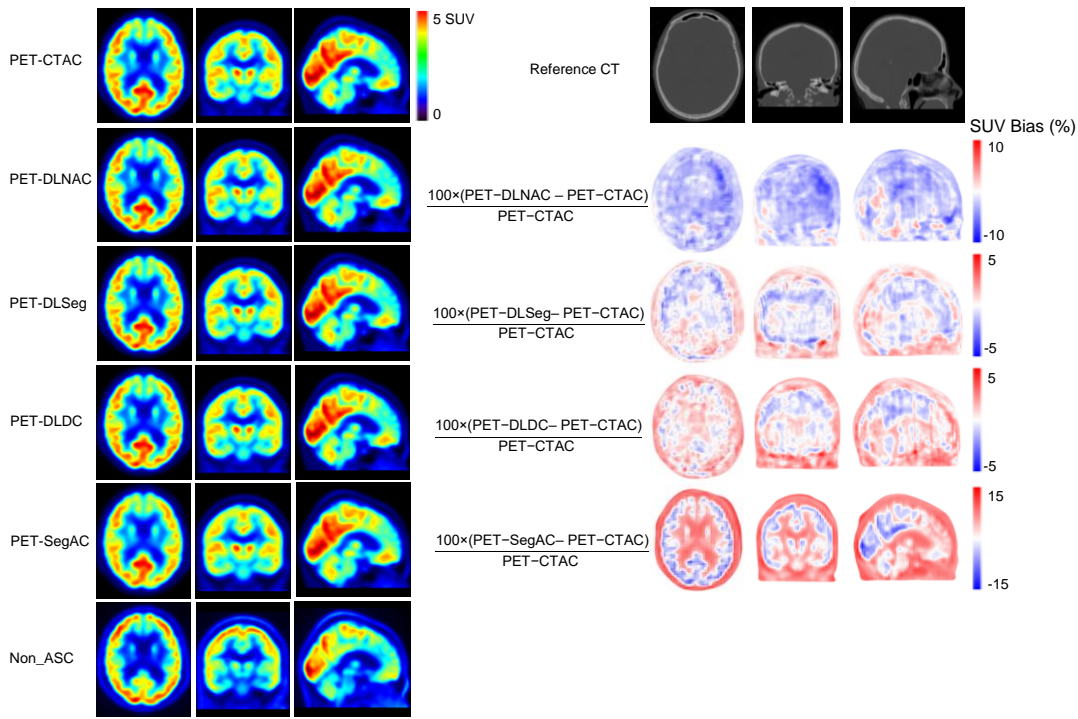

**Figure 1.** Representative transverse, coronal, and sagittal views of the PET-CTAC, PET-DLNAC, PET-DLSeg, PET-DLDC, and PET-SegAC together with their corresponding bias maps considering the PET-CTAC images as the reference.

The quantitative accuracy of different AC approaches was further investigated using a region-wise analysis of the brain PET images transformed into a common spatial map. Table 3 presents the mean SUV difference and bias calculated over 70 brain regions across 80 patients. Overall, the PET-DLDC exhibited a small SUV difference and bias across the entire brain regions.

**Table 3.** Mean SUV difference and relative SUV bias calculated over 70 brain regions (using a common spatial coordinate map) across 80 subjects for different AC approaches.

|               | Mean SUV difference | Relative bias %     |
| ------------- | ------------------- | ------------------- |
| **PET-DLNAC** | $0.04 \pm 0.29$     | $0.61 \pm 5.19$     |
| **PET-DLSeg** | $0.09 \pm 0.28$     | $1.42 \pm 5.08$     |
| **PET-DLDC**  | $0.01 \pm 0.21$     | $0.25 \pm 4.72$     |
| **PET-SegAC** | $-1.11 \pm 0.62$    | $-10.46 \pm 5.36$   |

The details of the region-wise analysis of the brain PET images are reflected in Figures 2 and 3 wherein the mean SUV bias and the mean absolute SUV bias are reported for each of the 70 brain regions, separately for PET-SegAC, PET-DLNAC, PET-DLSeg, and PET-DLDC images. In these figures, the mean SUV biases are reported for all 80 patients. The PET-Seg images exhibited the highest SUV bias for most regions while PET-DLDC images showed less than 3% absolute SUV bias across the entire regions.

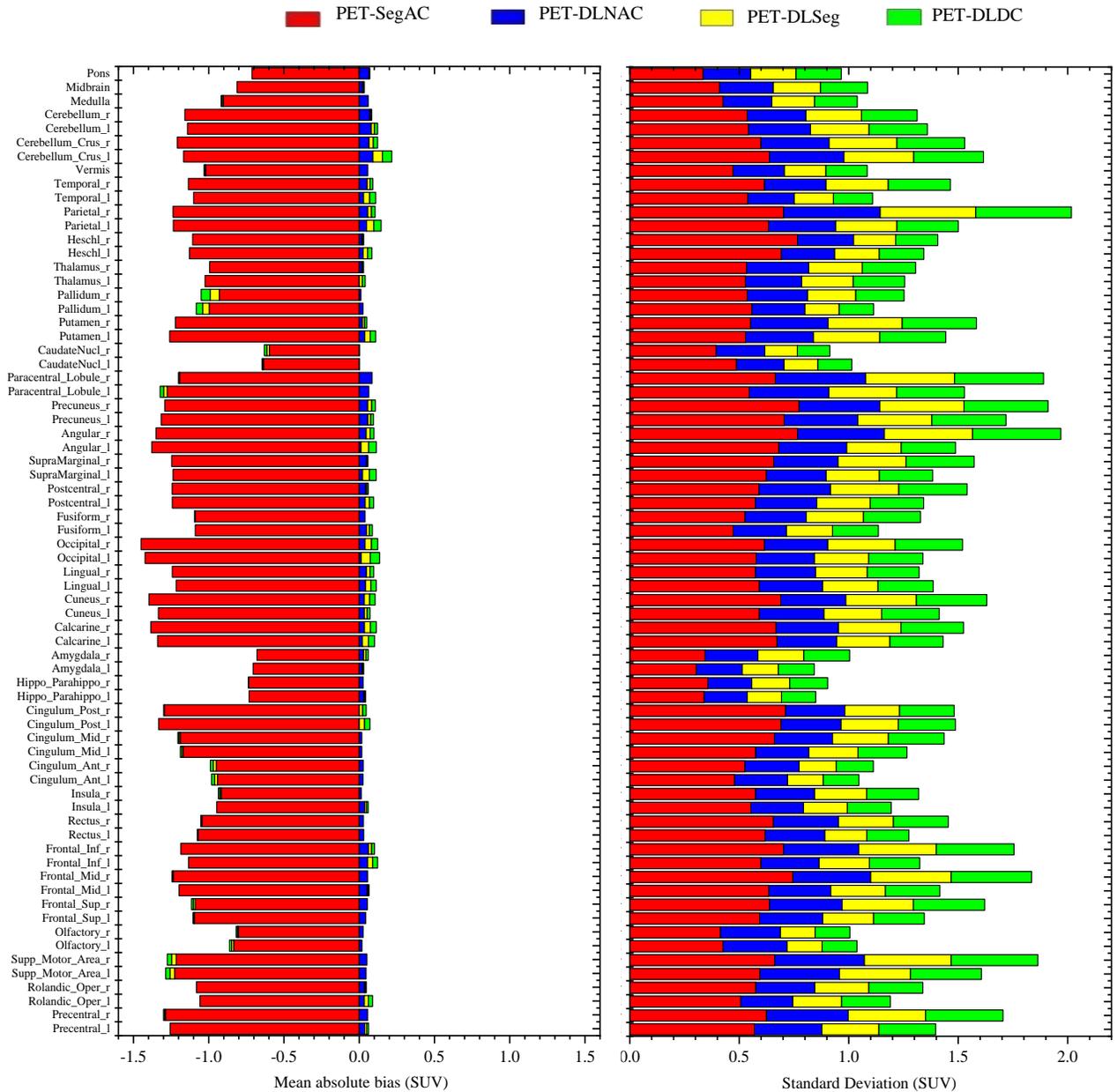

**Figure 2.** Mean SUV bias and standard deviation of the tracer uptake calculated in 70 brain regions for different PET images across 80 subjects.

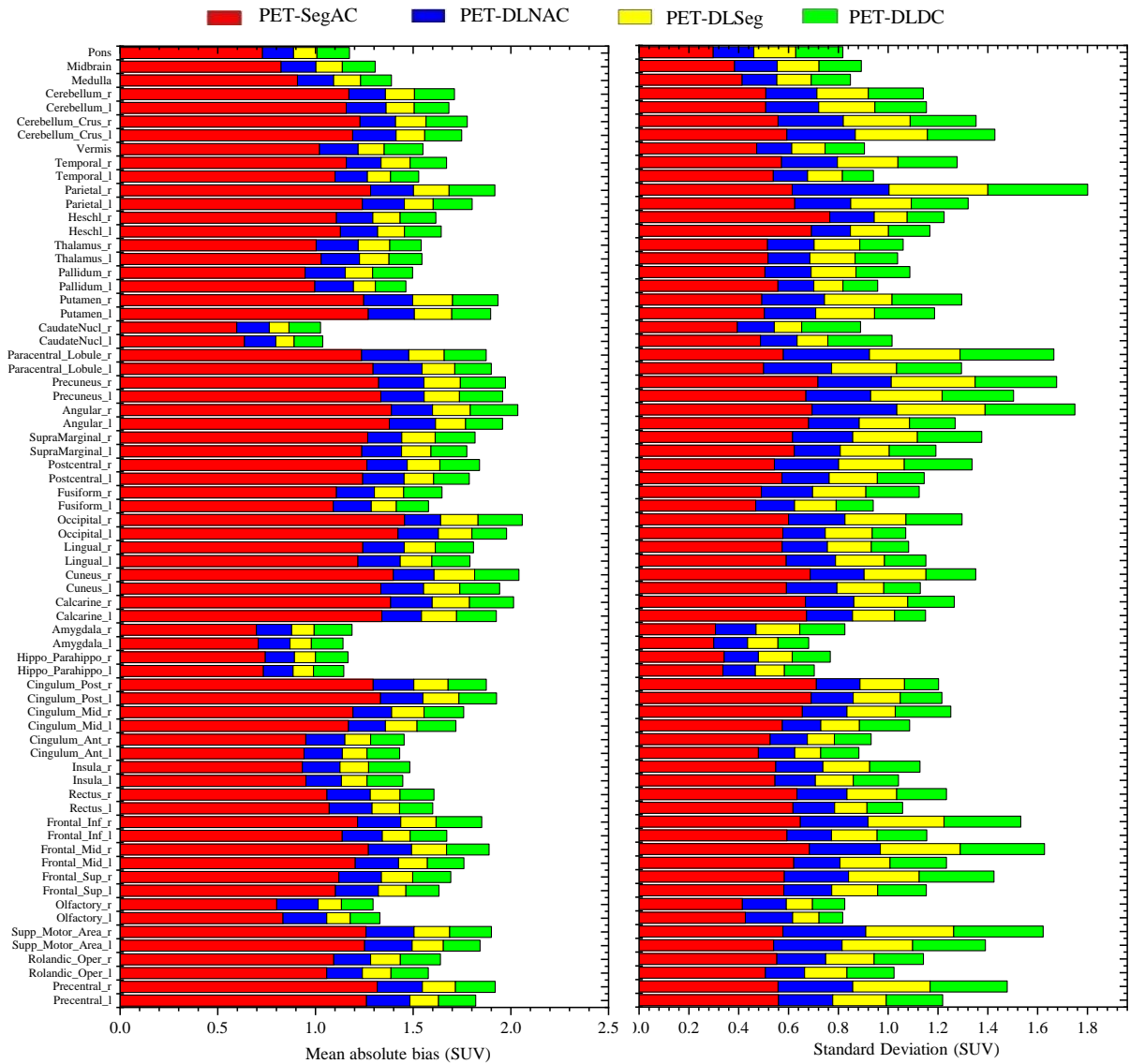

**Figure 3.** Mean absolute SUV bias and standard deviation of the tracer uptake calculated in 70 brain regions for different PET images across 80 subjects.

In total, 28 major radiomic features were calculated for each of the 70 brain regions and the different PET images. The relative errors of these features were calculated concerning the reference radiomic features obtained from PET-CTAC images. Figures 4 to 7 depict the relative errors of the 28 radiomic features for PET-SegAC, PET-DLNAC, PET-DLSeg, and PET-DLDC images.

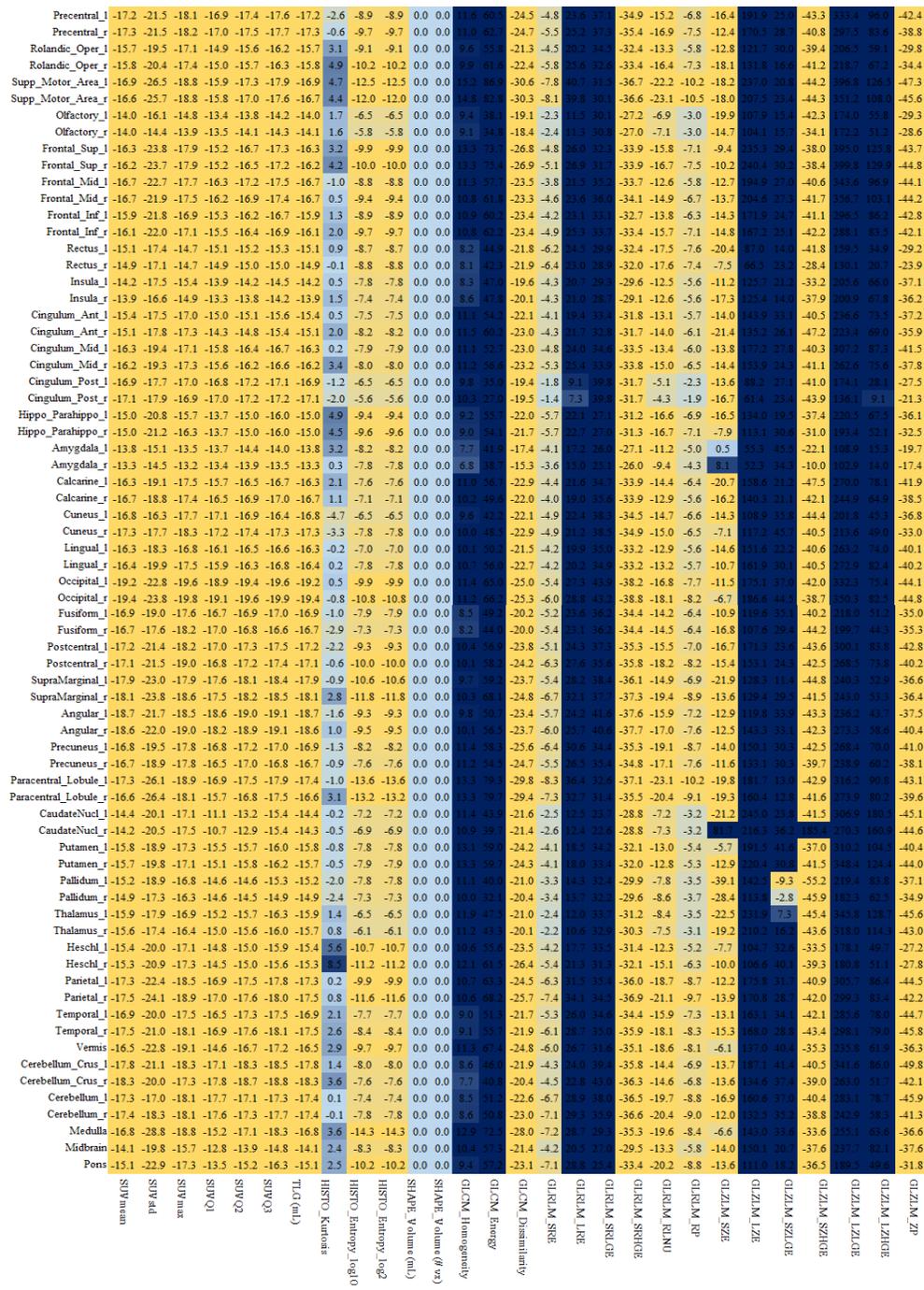

**Figure 4.** Heat map of the relative errors for the 28 radiomic features calculated across 70 brain regions in PET-SegAC images for the reference PET-CTAC images.

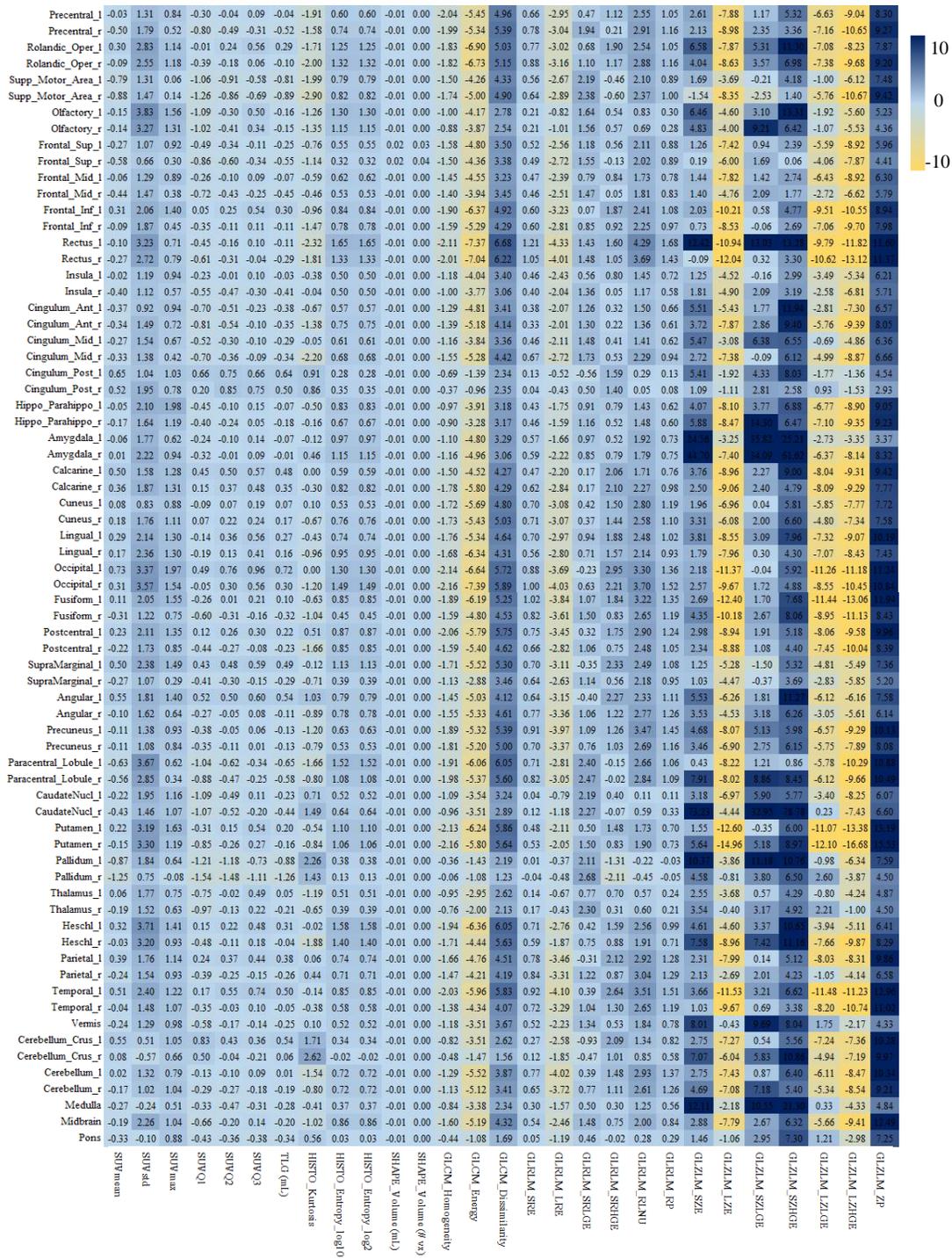

**Figure 5.** Heat map of the relative errors for the 28 radiomic features calculated across 70 brain regions in PET-DLNAC images for the reference PET-CTAC images.

**Figure 6.** Heat map of the relative errors for the 28 radiomic features calculated across 70 brain regions in PET-DLSeg images for the reference PET-CTAC images.

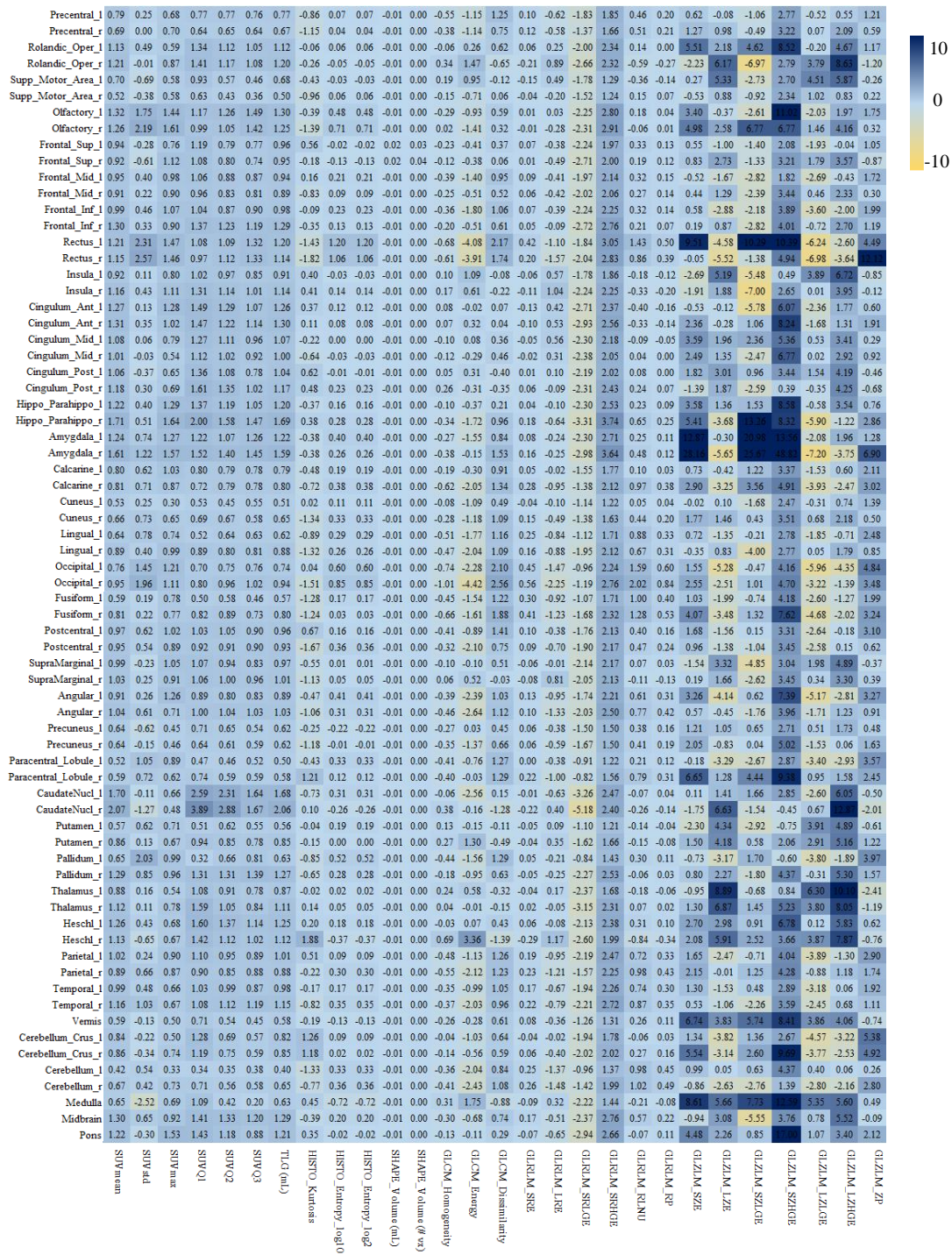

**Figure 7.** Heat map of the relative errors for the 28 radiomic features calculated across 70 brain regions in PET-DLDC images for the reference PET-CTAC images.

Quantitative analysis of the radiotracer uptake in the 70 brain regions for different PET images is summarized in Bland-Altman plots in Figure 8. Less bias and variance were observed in PET-DLDC images, compared to PET-DLSeg and PET-SegAC images; however, similar SUV bias and variance were seen in the PET-DLNAC images. The regression plots in Figure 9 and voxel-wise joint histogram analysis of the PET images revealed the superior accuracy of the radiotracer recovery in PET-DLDC images wherein RMSE=0.73 and $R^2$=1 were observed, compared to RMSE=1.29 and $R^2$=0.977, RMSE=1.87 and $R^2$=-0.999, and RMSE=1.10 and $R^2$=0.986 obtained from PET-SegAC, PET-DLSeg, and PET-DLNAC images, respectively.

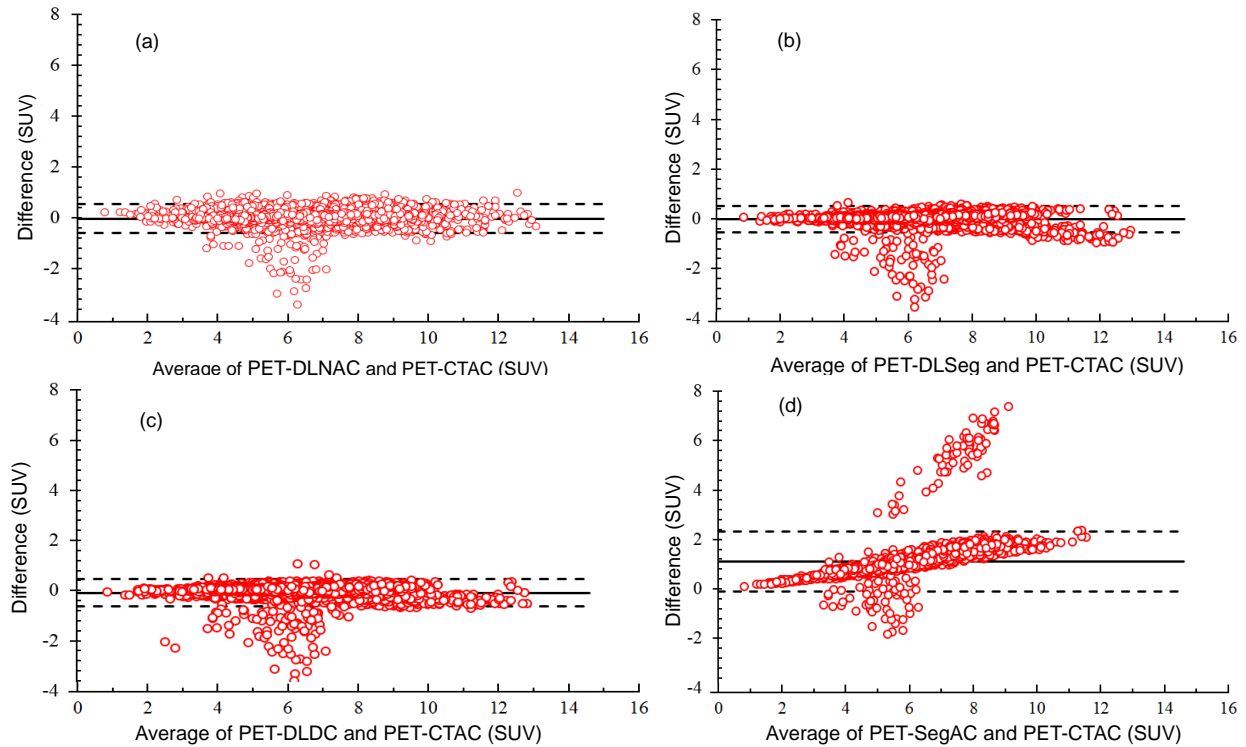

**Figure 8.** Bland-Altman plots of radiotracer uptake (SUV) in 70 brain regions estimated on the (a) PET-SegAC, (b) PET-DLNAC, (c) PET-DLDC, and (d) PET-DLSeg images for the reference PET-CTAC images. The solid line indicates the mean SUV difference, and the dashed lines show a 95% confidence interval (CI) of the SUV differences.

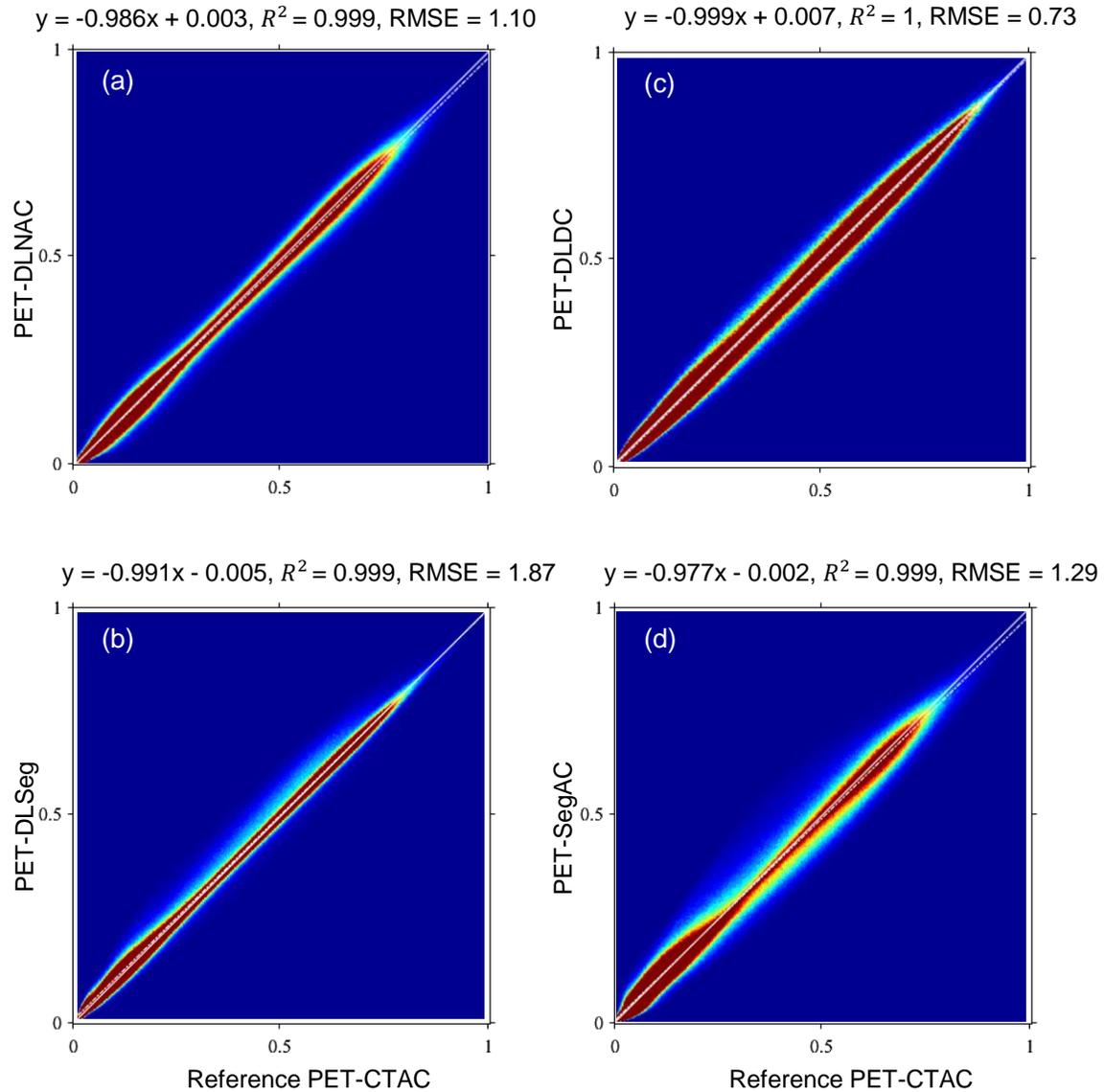

**Figure 9.** Joint histogram analysis of (a) PET-DLNAC, (b) PET-DLSeg, (c) PET-DLDC, and (d) PET-SegAC images versus the reference PET-CTAC images.

## 4. Discussion

Deep learning-based attenuation correction in PET imaging has recently attracted much attention by proposing various strategies of attenuation map generation or/and AC application [13, 14, 31-33]. Among these strategies, the direct application of the AC on PET-nonAC images in the image domain is of spatial interest. This approach does not require anatomical imaging, which is appealing/suitable for dedicated brain and pelvis PET scanners, and this approach is less sensitive to some common sources of error in conventional PET attenuation correction, such as mismatches between emission and transmission/MR images, body truncation, and inaccurate scatter correction [3, 19, 21, 22, 34]. In this regard, this study set out to investigate the use of a simple version attenuation and scatter corrected PET images in order to aid the direct application of AC in the image domain. The motivation behind this investigation was that a simple 2-class attenuation map can be easily generated from PET-nonAC images to reconstruct PET-Seg images. This auxiliary attenuation map would be in perfect alignment with the emission data since it is generated directly from the PET data.

The simple 2-class attenuation map, providing a bottom line for the comparison of different methods, was included in this study to highlight the levels of accuracy brought by the direct AC in the image domain. All of the three deep learning-based AC approaches evaluated in this study noticeably outperformed the segmentation-based AC map. The deep learning-based methods exhibited very close attenuation correction performances; however, in general, the dual-channel deep learning method taking both PET-nonAC and PET-SegAC images as the input showed slightly superior accuracy leading to an overall lower RMSE in the whole brain (Table 1) and the voxel-wise analysis (Figure 9). The differences were statistically significant between the dual-channel deep learning-based method and other methods for key parameters, such as the RMSE and the SUV bias in brain regions.

The 2-class AC map is far from perfect; nevertheless, it could help to partly compensate for the impact of attenuated and scattered photon on PET images. Therefore, the estimation of the PET-CTAC from the PET-SegAC would face less complexity. However, relying only on PET-SegAC images has the drawback of losing the original information that lies in PET-nonAC images. In this regard, any errors in PET-SegAC images, caused by miss-segmentation, erroneous AC map, or incorrect scatter correction would be reflected in the output of the deep learning method relying only on PET-SegAC images. A dual-channel deep learning method could benefit from the information existing in the combination of PET-nonAC and PET-SegAC images. In this sense, the PET-SegAC image helps to reduce the complexity of the AC problem and still the information which lies in the original PET-nonAC images would aid the deep learning approach to find an optimal solution. Moreover, in case of any error in PET-SegAC images, for example, due to the miss-segmentation, PET-nonAC images have sufficient information for the prediction of an acceptable/tolerable PET-CTAC image.

A significant/important component of the AC process is the correction of the scattered photon. Due to the model-based and analytic nature of the scatter estimation, the scatter distribution and factor/amplitude depend on the algorithm and parameters used for the calculation of the scattered photons. In this regard, the training dataset for direct AC in the image domain should be created using the same scatter correction algorithm and parameters to avoid discrepancies between different training subjects. Moreover, the same scatter correction setting should be employed for PET-SegAC images (to be used in the dual-channel deep learning approach) to create a consistent model. Previous studies have shown that scattered event estimation has a negligible sensitivity to attenuation maps wherein the scatter estimations from the reference CT image and a segmentation-based method, such as a 2-class AC map, are in close agreement [18, 35]. In this regard, since PET-SegAC images contain almost complete/sufficient information about the scattered events, the dual-channel deep learning model would be able to offer an overall superior solution due to the reduced complexity of the AC problem.

## 5. Conclusion

This study set out to investigate the direct application of AC in the image domain. To this end, three different scenarios were followed to train a deep learning model for the direct AC wherein the input of the network was assigned to only PET-nonAC image, only PET-SegAC (using a 2-class segmentation-based AC map obtained from the PET-nonAC), and both PET-nonAC and PET-SegAC images in a dual-channel mode. Overall, the dual-channel deep learning model taking both PET-nonAC and PET-SegAC images as the input exhibited superior accuracy for radiotracer recovery in brain imaging. Less than 4% SUV bias was observed in different regions of the brain when using the dual-channel deep learning model, while the segmentation-based AC map led to more than 13% SUV bias. The dual-channel deep learning model, which relies on both PET-nonAC and PET-SegAC images, would offer more accurate AC modeling, compared to a model using only PET-nonAC or PET-SegAC images.